\documentclass{article}
\usepackage{epsfig,multicol,cite}
\begin{document}
\parindent=0.2in

\newcommand{\be}{\begin{equation}}
\newcommand{\ee}{\end{equation}}
\newcommand{\beq}{\begin{eqnarray}}
\newcommand{\eeq}{\end{eqnarray}}
\newcommand{\ds}{\displaystyle}
\newcommand{\pia}{\mbox{$p_i^{\alpha}$}}
\newcommand{\pjb}{\mbox{$p_j^{\beta}$}}
\newcommand{\la}{\lambda_{\alpha}}
\newcommand{\bla}{\bar{\lambda}_{\alpha}}
\newcommand{\xa}{x^{\alpha}}
\newcommand{\ya}{y^{\alpha}}

\noindent
{
\bfseries\LARGE 
A model for the emergence of cooperation, interdependence and structure
in evolving networks} \\ \\
{\bf {Sanjay Jain$^{*\dag\ddag 1}$ and Sandeep Krishna$^{\S 2}$}}\\ \\
$^*$ Centre for Theoretical Studies and $^{\S}$ Physics Department,
Indian Institute of Science,\\ \hspace*{0.2in} Bangalore 560 012,
India\\
$^{\dag}$ Santa Fe Institute, 1399 Hyde Park Road, Santa Fe, NM 87501, USA\\
$^{\ddag}$ Jawaharlal Nehru Centre for Advanced Scientific Research,
Bangalore 560 064, India\\
$^1$ {\it jain@santafe.edu, jain@cts.iisc.ernet.in}; 
$^2$ {\it sandeep@physics.iisc.ernet.in} 

\begin{multicols}{2} 
\noindent
{\bf Evolution produces complex and structured networks of interacting 
components in chemical, biological, and social
systems.
We describe a simple mathematical model for the evolution of an idealized chemical
system to study how a network of cooperative molecular species arises and
evolves to become more complex and structured.
The network is modeled by a directed weighted
graph whose positive and negative links represent `catalytic' and `inhibitory'
interactions among the molecular species, and which evolves as the least populated 
species (typically those that go extinct) are replaced by new ones.
A small autocatalytic set (ACS),
appearing by chance, provides the seed for
the spontaneous growth of connectivity and cooperation in the graph. 
A highly structured chemical organization arises inevitably 
as the ACS enlarges and percolates through the network in a short,
analytically determined time scale. This self-organization does not
require the presence of self-replicating species. The network also exhibits
catastrophes over long time scales triggered by the chance 
elimination of `keystone' species, followed by recoveries.}

\vspace{0.1in}
Structured networks of interacting components are a hallmark of
several complex systems, for example, 
the chemical network of molecular species in cells \cite{Darnelletal}, the web of
interdependent biological species in ecosystems \cite{CBN,Pimm}, 
and social and economic networks
of interacting agents in societies \cite{WB,Axelrod,WS,BL}. 
The structure of these networks is a product of
evolution, shaped partly by the environment and physical constraints and
partly by the population (or other) dynamics in the system. For example, imagine a pond on
the prebiotic earth containing a set of interacting molecular species
with some concentrations. The interactions 
among the species in the pond affect how the populations evolve with time.
If some population goes to zero, or if new molecular species enter
the pond from the environment (through storms, floods or tides), the 
effective chemical network existing in the pond changes.
We discuss a mathematical model that attempts to incorporate this interplay
between a network, populations, and the environment in a simple and
idealized fashion. The model (including an earlier version \cite{JK1,JK2}) was inspired
by the ideas and results in refs. 
\cite{Dyson,Kauffman2,Kauffman3,FKP,BF,BFF,Fontana,FB,BS}. 
Related but different models are studied in refs. \cite{HS,YT,SBL}.

\vspace{0.2in}
\noindent {\large\bf The Model}

\vspace{0.1in} \noindent
The system consists of $s$ species labeled by the index $i=1,2, \ldots, s$.
The network of interactions between species is specified
by the $s \times s$ real matrix  $C \equiv \{c_{ij}\}$. 
The network can be visualised as a directed
graph whose nodes represent the species. A non-zero $c_{ij}$ is represented by
a directed weighted link from node $j$ to node $i$. If
$c_{ij} > 0$ then the corresponding link is a cooperative 
link: species $j$ catalyzes the production of species $i$. 
If $c_{ij} < 0$ it is a destructive link: the presence of $j$ causes a depletion of $i$.

\vspace{0.1in} \noindent {\bf Population dynamics.} The model contains
another dynamical variable  
${\bf x}\equiv (x_1, \ldots. x_s)$, where $x_i$ stands for
the relative population of the $i^{th}$  species ($0 \leq x_i \leq 1$,
$\sum_{i=1}^s x_i = 1$). The time evolution of ${\bf x}$ depends upon the
interaction coefficients $C$, as is usual in population models. The specific
evolution rule we consider is 
\be
\label{xdot}
\begin{tabular}{llllcl}
$\dot{x}_i$ & $=$ & $f_i$ & \hspace*{0.1cm} if $x_i>0$ & or & $f_i\geq 0$,\\
            & $=$ & $0$   & \hspace*{0.1cm} if $x_i=0$ & and & $f_i<0$,\\
\end{tabular}\\
\ee
\vspace*{-0.5cm}
$$\mbox{where } f_i=\sum_{j=1}^{s}c_{ij}x_j-x_i\sum_{k,j=1}^{s}c_{kj}x_j.$$
This is a particularly simple idealization of catalyzed chemical reaction
dynamics in a well stirred reactor, as explained later.

\vspace{0.1in} \noindent {\bf Graph dynamics.} The dynamics of 
$C$ in turn depends upon ${\bf x}$, as follows:  
Start with a random graph of $s$ nodes: $c_{ij}$ is non-zero with probability $p$
and zero with probability $1-p$. If nonzero, $c_{ij}$ is chosen randomly in the interval 
$[-1,1]$ for $i\ne j$ and $[-1,0]$ for $i=j$. Thus a link
between two distinct species, if it exists, 
is just as likely to be cooperative as destructive, 
and a link from a species to itself can only be inhibitive, i.e., autocatalytic or 
self-replicating individual species are not allowed. 
The variable ${\bf x}$ is initialized
by choosing each $x_i$ randomly between $0$ and $1$, and then rescaling all
$x_i$ uniformly such that $\sum_{i=1}^s x_i = 1$.
The evolution of the network proceeds 
in three steps : \\
1) Keeping the network fixed, the populations are evolved according to (\ref{xdot})
for a time $T$ which is large enough for ${\bf x}$ to get
reasonably close to its attractor. We denote $X_i \equiv x_i(T)$. \\
2) The set of nodes $i$ with the least value of $X_i$ is determined. 
We call this the set of `least fit' nodes, identifying the relative population of a species
in the attractor (or, more specifically, at $T$)
with its `fitness' in the environment defined by the graph.
One of the least fit nodes is chosen randomly (say $i_0$) and removed from the system
along with all its links leaving a graph of $s-1$ species.\\
3) A new node is added to the graph so that it again has $s$ nodes. The 
links of the added node ($c_{ii_0}$ and $c_{i_0i}$, 
for $i=1, \ldots, s$) are assigned
randomly according to the same rule as for the nodes in the initial graph.
The new species is given a small relative population $x_{i_0}=x_0$ and the
other populations are rescaled to keep $\sum_{i=1}^s x_i=1$. This process,
from step 1 onwards, is iterated many times.

\vspace{0.1in} \noindent {\bf Motivation for model structure.}
The choice of Eq. (\ref{xdot}) is motivated by the rate equations in a 
well stirred chemical reactor (representing, say, a prebiotic pond) as follows:
If species $j$  catalyses the ligation of reactants $A$ and $B$ to form the species $i$,
$A + B\stackrel{j}{\rightarrow} i$, then the rate of growth of the population
$y_i$ of species $i$  will be given by
$\dot{y}_i=k(1+\nu y_j)n_An_B-\phi y_i$,
where $n_A, n_B$ are reactant concentrations,
$k$ is the rate constant for the spontaneous reaction,
$\nu$ is the catalytic efficiency, and $\phi$ represents a common
death rate or dilution flux in the reactor \cite{Ashmore}.
Assuming the catalysed reaction is much faster than the spontaneous
reaction, and the concentrations of the reactants
are large and fixed, the rate equation becomes
$\dot{y}_i=cy_j-\phi y_i$, where $c$ is a constant. If species $i$ has multiple
catalysts,  we get $\dot{y}_i = \sum_{j=1}^{s}c_{ij}y_j-\phi y_i$. The
first of equations (\ref{xdot}) follows from this upon using
the definition $x_i=y_i/ \sum_{j=1}^s y_j$.
When negative links are permitted, the second of equations (\ref{xdot}) is
needed in general to prevent relative populations from going negative.
(With negative links, a more realistic chemical interpretation would be
obtained if $\dot{x}_i$ were proportional to $x_i$, but for simplicity we
retain the form of (\ref{xdot}) in this paper.) (\ref{xdot}) may be
viewed as defining an artificial chemistry in the spirit of
refs. \cite{FKP,BF,BFF,Fontana,FB}.

The rules for the evolution of the network $C$ are intended to capture two key features of natural evolution, namely, 
selection and novelty. The species that has the least population in the
attractor configuration is the one most likely to be eliminated in a large fluctuation
in a possible hostile environment. Often, the least
value of $X_i$ is zero. Thus the model implements selection by
eliminating from the network a species that has become extinct or has the
least chance of survival \cite{BS}. 
Novelty is introduced in the network in the form of a new species. This species
has on average the same connectivity as the initial set of species, but its actual
connections with the existing set are drawn randomly. 
E.g., if a storm brings into a prebiotic pond a new molecular species from the
environment, the new species might be statistically similar to the one being eliminated,
but its actual catalytic and inhibitory interactions with the surviving species
can be quite different. Another common feature of
natural evolution is that populations typically evolve on a fast time scale compared
to the network. This is captured in the model by having the $x_i$ relax to 
their attractor before the network is updated.  The idealization of a fixed total 
number of species $s$ is one that we hope to relax in future work.

The model described above differs from the one studied in \cite{JK1,JK2}
in that it allows negative links and varying link strengths, and that the
population dynamics, given by (\ref{xdot}), is no longer linear.
The earlier model had only fixed point attractors; here
limit cycles are also observed. Since $C$ now has negative entries, 
the formalism of non-negative matrices no longer applies.

\vspace{0.2in} \noindent {\large \bf Results}

\vspace{0.1in} \noindent {\bf Emergence of cooperation and interdependence.}
Figure 1 shows a sample run. 
The same qualitative behaviour is seen in each of 
several hundred runs performed for $p$ values ranging from $0.00002$ to 
$0.01$ and for $s=100, 150, 200$. 
The fact that the ratio of number of cooperative to destructive links at first remains
constant at unity (statistically) and then increases 
by more than an order of magnitude is evidence of
the emergence of cooperation. The increase in $\bar{d}$ 
by an order of magnitude is a quantitative 
measure of the increase of interdependence of species in the network. 
The increase in the total density of links $(l_+ + l_-)/s$ is another aspect
of the increase of complexity of the system. 
Note that in the model selection rewards
only `performance' as measured in terms of relative population; the rules do not 
select for higher cooperativity per se. Since a new species is equally
likely to have positive or negative links with other species, the introduction of
novelty is also not biased in favour of cooperativity. 
The fact that this behaviour is not a consequence of any intrinsic bias in 
the model that favours the increase of cooperation and interdependence
is evidenced by the flat initial region of all the curves.

\vspace{0.1in} \noindent {\bf Autocatalytic sets.}
The explanation for the above behaviour lies in the formation and growth of 
certain structures, autocatalytic sets (ACSs), in the graph.
An ACS is defined as a set of nodes such that each node has at least one
incoming positive link from a node in the set. Thus an ACS has the property
of catalytic closure, i.e., it contains a catalyst for each of its 
members \cite{Eigen,Rossler,Kauffman1}.
The simplest example of an ACS is a cycle of positive links. 
Every ACS is not such a cycle but it can be shown that an ACS must contain 
a cycle of positive links \cite{JK2}. In Fig. 1, there is no ACS in the
graph until $n=1903$. A small ACS (which happens to be a cycle of positive
links between two nodes) appears at $n \equiv n_1 = 1904$, exactly
where the behaviour of the $s_1$ curve changes. As time proceeds this ACS becomes 
more complex and enlarges until at $n \equiv n_2 = 3643$ the entire
graph becomes an ACS. $l_+$ and $\bar{d}$ exhibit an increase and $l_-$ a 
decrease as the ACS comes to occupy a significant part of the graph.
After the ACS first appears (at $n=n_1$), the set of populated nodes
in the attractor configuration ($s_1$ in number), is always an ACS
(except for certain catastrophic events to be discussed later),  
which we call the 
`dominant ACS'. The spontaneous appearance of a small ACS at some $n=n_1$, 
its persistence (except for catastrophes), and its
growth until it spans the graph at $n=n_2$, is seen in each of the several hundred runs mentioned earlier.
The growth of the ACS across the graph between $n_1$ and $n_2$ occurs exponentially 
(with stochastic fluctuations),
\be \label{growth}
s_1(n)\approx s_1(n_1)e^{(n-n_1)/\tau_g}, \quad \quad \tau_g = 2/p. 
\ee
This agrees with simulations as shown in Figure 2. The average time scale $\tau_a \equiv 
\langle n_1 \rangle$ for the first appearance of the ACS is given, for 
sufficiently small $p$, by $\tau_a\approx 4/(p^2s)$ ($=1600$ for $p=0.005$ and 
$s=100$).

Upto $n=n_1$, the graph has no ACS. It has chains and trees of positive
and negative links and possibly loops containing negative links. These
latter structures are not robust. For example, consider
a chain of two positive links $1 \rightarrow 2 \rightarrow 3$. Since catalytic
links are pointing to node $3$, it will do well populationally compared to 
nodes $1$ and $2$. However, since $1$ has no incoming catalytic links, its
relative population will decline to zero under (\ref{xdot}), and it can be picked
for replacement in the next graph update. This can disrupt the chain and
hence erode the `well being' of node $3$ until eventually after some graph
updates the latter can also join the ranks of the least fit. 
Species $3$ gets eliminated eventually because it
does not feedback into and `protect' species $1$ and $2$, on whom its `well being'
depends.
In a graph without an ACS no structure is protected from disruption.
Since every node is liable to be replaced sooner or later, 
the graph remains as random as the initial graph 
(we have checked that the probability distribution of the number of incoming and 
outgoing links at a node remains the appropriate binomial for $n < n_1$).
This explains why $s_1$, $l_{\pm}$ and $\bar{d}$
hover around their initial values.

The picture changes the moment a small ACS appears in the graph. The key
point is that by virtue of catalytic closure, members of the ACS do well
{\it collectively} in the population dynamics governed by (\ref{xdot}). 
An ACS is a collective self-replicator and beats chains, trees and
other non ACS structures in the population game, reducing their $X_i$  to 
zero when it appears. Thus, since graph update proceeds by replacing
one of the nodes with $X_i = 0$ (if present) with a new one, such a replacement being 
outside the dominant ACS can cause no damage to the links that constitute 
the ACS. That is why the ACS structure, once it appears, is much more robust 
than the non-ACS structures discussed earlier. If the new node 
happens to get an incoming
positive link from the dominant ACS, it becomes part of it. Thus the
dominant ACS tends to expand in the graph as new nodes get attached to 
it \cite{BFF} \cite{JK1,JK2} and $s_1$ increases.  
In $\Delta n$ graph updates the average increase in $s_1$, which is 
the number of added nodes which will get a 
positive link from one of the $s_1$ nodes of the dominant ACS, is 
$\Delta s_1\approx (p/2)s_1\Delta n$, for small $p$. This proves (\ref{growth}). (Note that the exponential growth described by (\ref{growth}) is not
to be confused with the exponential growth of populations $y_i$ of species
that are part of the ACS. (\ref{growth}) reflects the growth of the 
ACS across the graph, or the increase in the number of species that
constitute the ACS.)

Since the dominant ACS grows by 
adding positive links from the existing dominant ACS, the number
of positive links increases as the ACS grows. On the other hand 
nodes receiving negative links usually end up being least fit, hence 
negative links get removed when these nodes are eliminated. 
Which novelty is captured thus depends upon the existing `context';
the network evolves by preferentially capturing links and nodes that 
`latch on' cooperatively to the existing ACS and disregarding those that
do not. The `context' itself arises when the ACS structure first appears;
this event transforms the nature of network evolution from random to
`purposeful' (in this case directed towards increasing cooperation). 
Before the ACS appears nothing interesting happens even though
selection is operative (the least populated species are being eliminated).
It is only after the ACS topological structure appears that selection 
for cooperation and complexity begins.

\vspace{0.1in} \noindent {\bf Inevitability of autocatalytic sets.}
Note that the appearance of an ACS, though a chance event, is inevitable.
For $sp  \ll 1$, the probability that a graph not containing a two cycle
will acquire one at the next time step is $p^2s/4 \equiv q$. Since the probability of 
occurrence of 3-cycles, etc., is much smaller, the probability 
distribution of arrival times $n_1$ is approximated by $P(n_1)= q(1-q)^{n_1 - 1}$, whose
mean $\tau_a$ is $1/q$. Since this probability declines exponentially after
a time scale $1/q$, the appearance of an ACS is inevitable, even for arbitrarily small
(but finite) $p$.

Occasionally in a graph update $s_1$ can decrease for various reasons.
If the new node forms an ACS of its own with nodes 
outside the dominant ACS, and the new ACS has a higher population growth rate 
(as determined by (\ref{xdot})) than the
old ACS it drives the species of the latter to extinction and becomes the
new dominant ACS.  Alternatively the new node could be a `destructive parasite':
it receives one or more positive links from and gives one or more negative links 
to  the dominant ACS. Then part or whole of the ACS 
may join the set of least fit nodes. Structures that diminish the size of the
dominant ACS or destroy it appear rarely. For example in Figure 1,
destructive parasites appeared $6$ times at $n=3388, 3478, 3576, 3579, 3592$ 
and $3613$. In each case $s_1$ decreased by 1.

\vspace{0.1in} \noindent {\bf Emergence of structure.}
At $n=n_2$ the whole graph becomes an ACS; the entire system can collectively
self-replicate despite the explicit absence of individual self-replicators.
Such a fully autocatalytic set is a very non-random structure.
Consider a graph of $s$ nodes and let the probability of a positive link 
existing between any pair of nodes be $p^*$. Such a graph has on average 
$m^*=p^*(s-1)$ incoming or outgoing positive links per node. 
For the entire graph to be an ACS, each node must have at least one
incoming positive link, i.e., each row of the matrix $C$ must contain at least one
positive element. Hence the probability, $P$, for the entire random graph to be an 
ACS is\\
\begin{tabular}{ccl}
$P$ & $=$ & probability that every row has at least \\  
    &     &  \quad \quad \quad \quad \quad \quad one positive entry\\
    & $=$ & [probability that a row has at least  \\
    &     &  \quad \quad \quad  \quad \quad \quad one positive entry$]^s$\\
    & $=$ & $[1-($probability that every entry \\ 
    &     & \quad \quad \quad \quad \quad \quad of a row is $\le 0)]^s$\\
    & $=$ & $[1-(1-p^*)^{s-1}]^s$\\
    & $=$ & $[1-(1-m^*/(s-1))^{s-1}]^s$.\\
\end{tabular}\\
For large $s$ and $m^*\sim O(1)$,
\be \label{prob} 
P\approx (1-e^{-m^*})^s=e^{-\alpha s}, \ee
where $\alpha$ is positive and $O(1)$. 
At $n=n_2$, we find in all our runs that $l_+(n_2) \equiv l^*$ 
is greater than $s$ but of order $s$, i.e., $m^* \sim O(1)$.
Thus dynamical evolution in the model via the ACS mechanism converts a
random organization into a highly structured one that is exponentially unlikely
to appear by chance. In the displayed run at $n=n_2$ the graph had $117$ positive links. 
The probability that a random graph with $s=100$ nodes and $m^*=1.17$ 
would be an ACS is given by (\ref{prob}) to be $\approx 10^{-16}$. 

Such a structure would take an exponentially long time to arise by pure
chance. The reason it arises inevitably in a short time scale in the 
present model is the following: a {\it small} ACS can appear by chance
quite readily, and once appeared, it grows exponentially fast across the
graph by the mechanism outlined earlier. 
The dynamical appearance of such a structure may be 
regarded as the emergence of `organizational
order'. The appearance of `exponentially unlikely' structures in the 
prebiotic context has been a puzzle. The fact that 
in the present model such structures inevitably form in a 
short time may be relevant for the origin of life problem.

\vspace{0.1in} \noindent {\bf The self-organization time scale in a prebiotic
scenario.} We now speculate on a possible application to prebiotic
chemical evolution.  
Imagine the molecular species to be small peptide
chains with weak catalytic activity in a prebiotic pond alluded to earlier. The pond
periodically receives an influx of new molecular species being randomly generated
elsewhere in the environment, through tides, storms or floods. Between these
influxes of novelty the pond behaves as a well stirred reactor where the 
populations of existent molecular species evolve according to 
(a realistic version of) eq. (\ref{xdot}) and reach their attractor configuration.
Under the assumption that the present model captures what happens in 
such a pond, the growth timescale (\ref{growth}) for a highly structured almost 
fully autocatalytic chemical organization in the pond is $\tau_g = 2/p$
in units of the graph update time step. In this scenario, the latter time unit corresponds
to the periodicity of the influx of new molecular species, hence it ranges from
one day (for tides) to one year (for floods). Further, in the present model
$p/2$ is the probability that 
a random small peptide will catalyse the production of 
another \cite{Kauffman1}, and this has 
been estimated in \cite{Kauffman3}: $p/2 \sim 10^{-5} - 10^{-10}$. 
With $p/2 \sim 10^{-8}$, 
for example, the time scale for a highly structured chemical organization to  
grow in the pond would be estimated to be of the order of $10^6$ to $10^8$ 
years. It is believed that life originated on Earth in a few hundred million
years after the oceans condensed.
These considerations suggest that it might be worthwhile to empirically pin down the `catalytic probability' $p$ (introduced in \cite{Kauffman1})
for peptides, catalytic RNA, lipids, etc., on the one hand, 
and explore chemically more realistic models on the other.

\vspace{0.1in} \noindent {\bf Catastrophes and recoveries in the network
dynamics.} 
After $n=n_2$ the character of the network evolution changes 
again. For the first time the least fit node will be one of the ACS members. 
Most of the time elimination of the node does not affect the ACS 
significantly and $s_1$ fluctuates between $s$ and $s-1$. 
Sometimes the least fit node could be a `keystone' species which
plays an important organizational role in the network despite its low 
population. When such a node is eliminated many other nodes can get disconnected
from the ACS resulting in large dips in $s_1$ and $\bar{d}$ and subsequently 
large fluctuations in $l_+$ and $l_-$. These large `extinction events' can be
seen in Figure 3.
Occasionally the ACS can even be destroyed completely. The system recovers
on the time scale $\tau_g$ after large extinctions 
if the ACS is not completely destroyed; if it is and the next few
updates obliterate the memory of previous structures in the graph, then
again a time on average $\tau_a$ elapses before an an ACS arises and the 
self-organization process begins anew. It may be of interest (especially in ecology,
economics, and finance), that network dynamics based on a fitness selection and
the `incremental' introduction of novelty, as discussed here, can by itself
cause catastrophic events without the presence of large external perturbations.

\vspace{0.2in} \noindent {\large \bf Discussion}

\vspace{0.1in} \noindent We have described an evolutionary model in which
the dynamics of species' populations (fast variables) and the graph of 
interactions among them (slow variables) are mutually coupled. The network
dynamics displays self-organization seeded by the chance but inevitable 
appearance of a small cooperative structure, namely, an ACS. In a dynamics
that penalizes species for low population performance, the collective 
cooperativity of the ACS members makes the set relatively robust against
disruption. New species that happen to latch on cooperatively to 
this structure preferentially survive, further enlarging the ACS in the
process. Eventually the graph acquires a highly non-random structure.
We have discussed the time evolution of quantitative measures of
cooperation, interdependence and structure of the network, which capture
various aspects of the complexity of the system.

It is noteworthy that
collectively replicating ACSs arise even though individual
species are not self-replicating. Thus the present mechanism is different from
the hypercycle \cite{ES}, where a template is needed to produce copies
of existing species. Unlike the hypercycle, the ACS is not disrupted by
parasites and short-circuits and grows in complexity, 
as evidenced in all our runs. It can be disrupted, however, when it
loses a `keystone' species.

It is also worth mentioning one departure from \cite{Kauffman3},
in that we find that a fully autocatalytic system (or percolating ACS) 
is not needed 
apriori for self-organization. In the present model a small ACS, once formed, 
typically expands (see also \cite{BFF}) and eventually
percolates the whole network dynamically. This dynamical process  
might be relevant for economic takeoff and technological growth in societies.

\vspace{0.2in}
\noindent
{\bf Acknowledgements.} We thank J. D. Farmer and W. C. Saslaw for helpful 
comments on the
manuscript. S. J. acknowledges the Associateship of Abdus Salam International
Centre for Theoretical Physics.
\vfill

\pagebreak

\vfill
\end{multicols}

\pagebreak 
\begin{figure}
\epsfxsize=8cm
\noindent
{\bf a\\}
\epsfbox{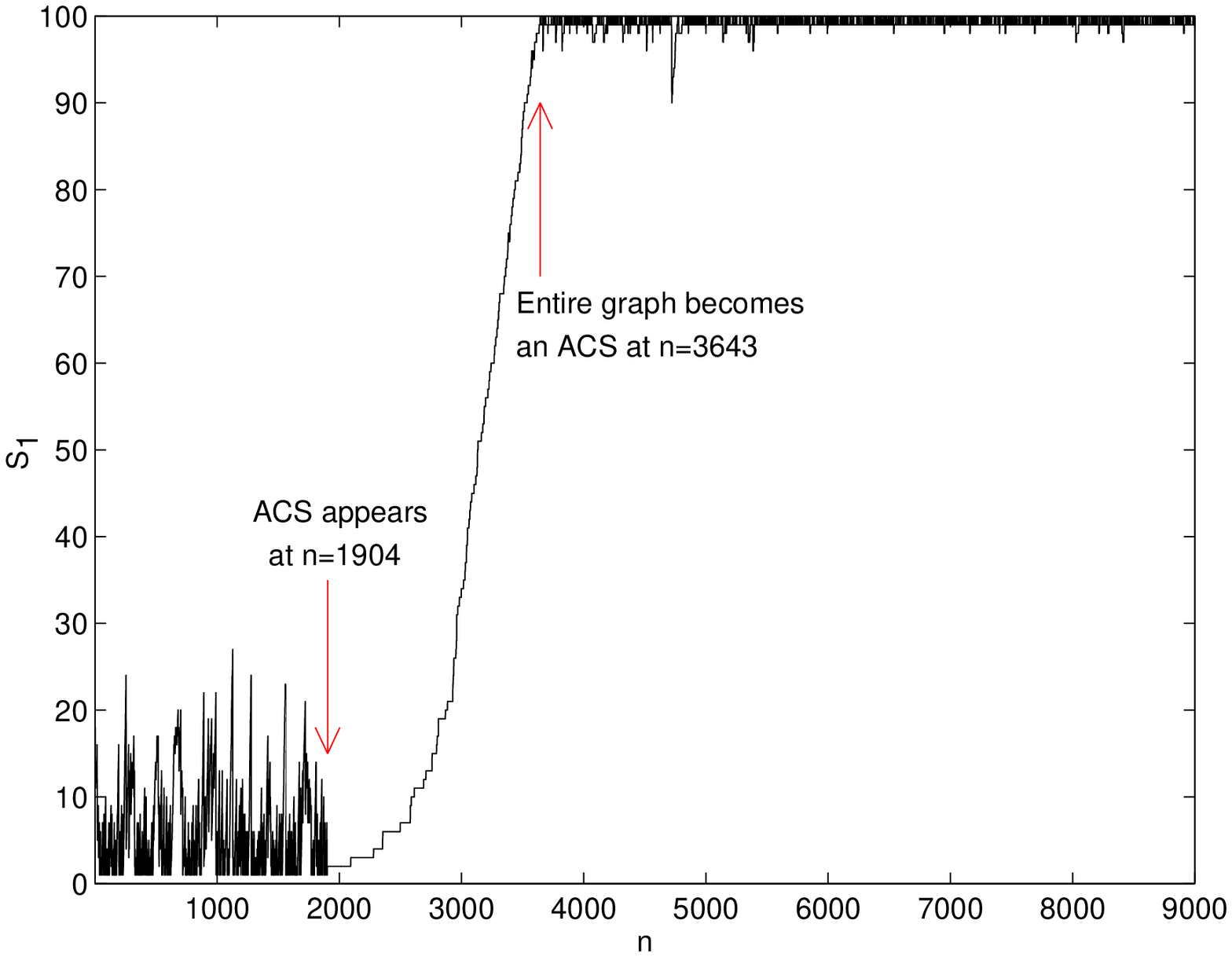}
\epsfxsize=8cm
{\bf \\ b\\}
\epsfbox{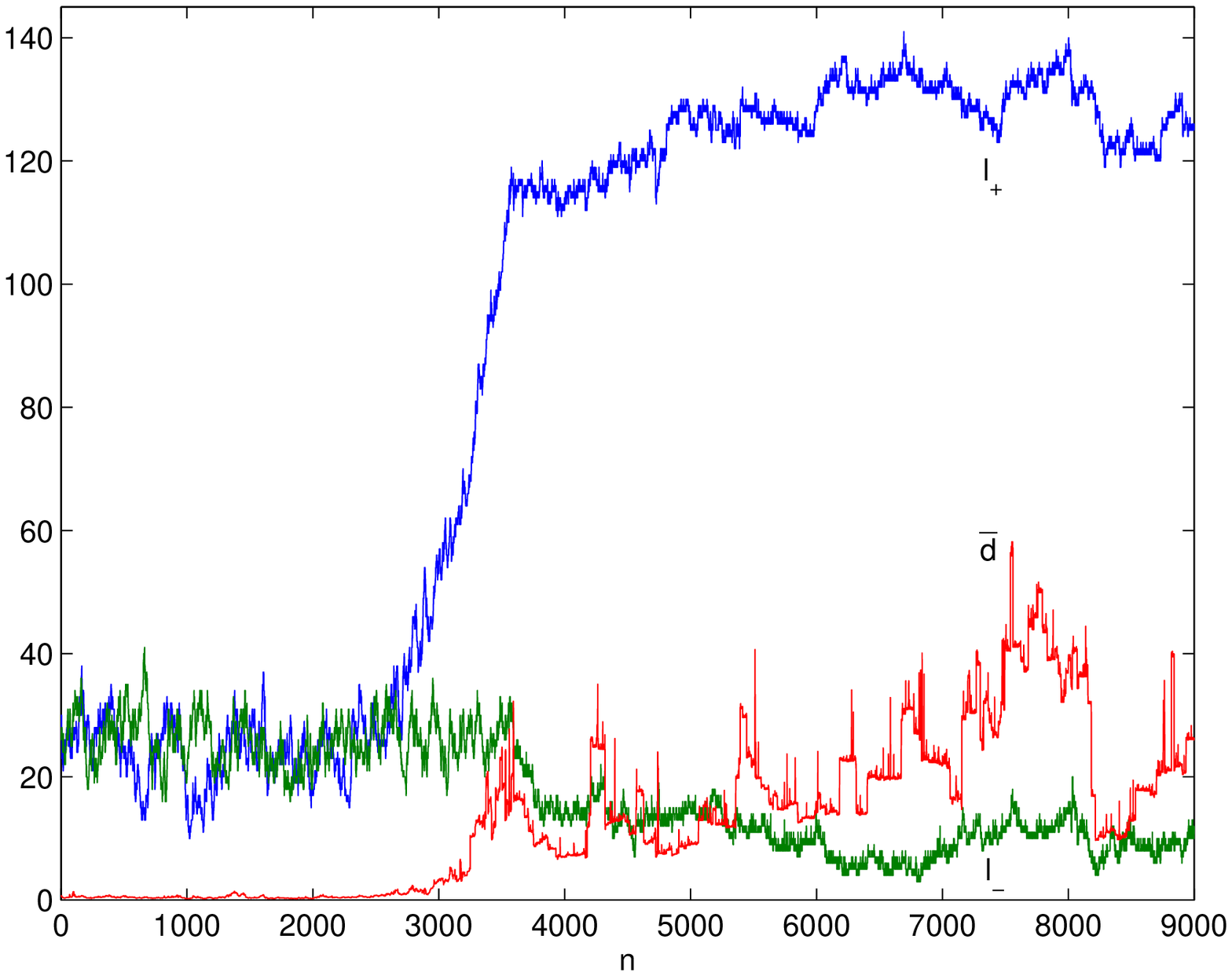} 

\vspace{0.1in}
\noindent
{\bf Figure 1.} A run with parameter values 
$s=100$, $p=0.005$, and $x_0=10^{-4}$.\\
({\bf a}) Number of populated species, $s_1$, in the attractor of
(\ref{xdot}) (i.e., number of nodes with $X_i > 0$)
after the $n^{th}$ addition of a new species (i.e., after $n$ graph update
time steps). ({\bf b}) The number, $l_+$, of positive links ($c_{ij}>0$) in the graph (blue); 
the number, $l_-$, of negative links (green); and `interdependency', $\bar{d}$, 
of the species in the network (red). Interdependency is defined as
$\bar{d}\equiv(1/s)\sum_{i=1}^s d_i$, where $d_i$ is the `dependency' 
of the $i^{th}$ node. $d_i \equiv \sum_{kj} |c_{kj}|h^i_k$, where 
$h^i_k$ is $1$ if there exists a directed path from $k$ to $i$ and 
$0$ otherwise. $d_i$ is the sum of (the absolute value of) the strengths
of all links that eventually feed into  $i$ along some directed path.
The curves have three distinct regions. Initially $s_1$ is small;
most of the species have zero relative populations. $l_+$ and $l_-$ also
do not vary much from their initial (random graph) value ($\approx ps^2/2=25$) 
and remain approximately equal. $\bar{d}$ hovers about its initial low value.
In the second region $s_1$, $l_+$ and $\bar{d}$ show a sharp increase 
and $l_-$ decreases. In the third region 
$s_1$, $l_{\pm}$ and $\bar{d}$ level off (but with fluctuations) and almost all species 
have non-zero populations in contrast to the initial period. \\

\end{figure}

\begin{figure}
\epsfxsize=8cm
\noindent
\epsfbox{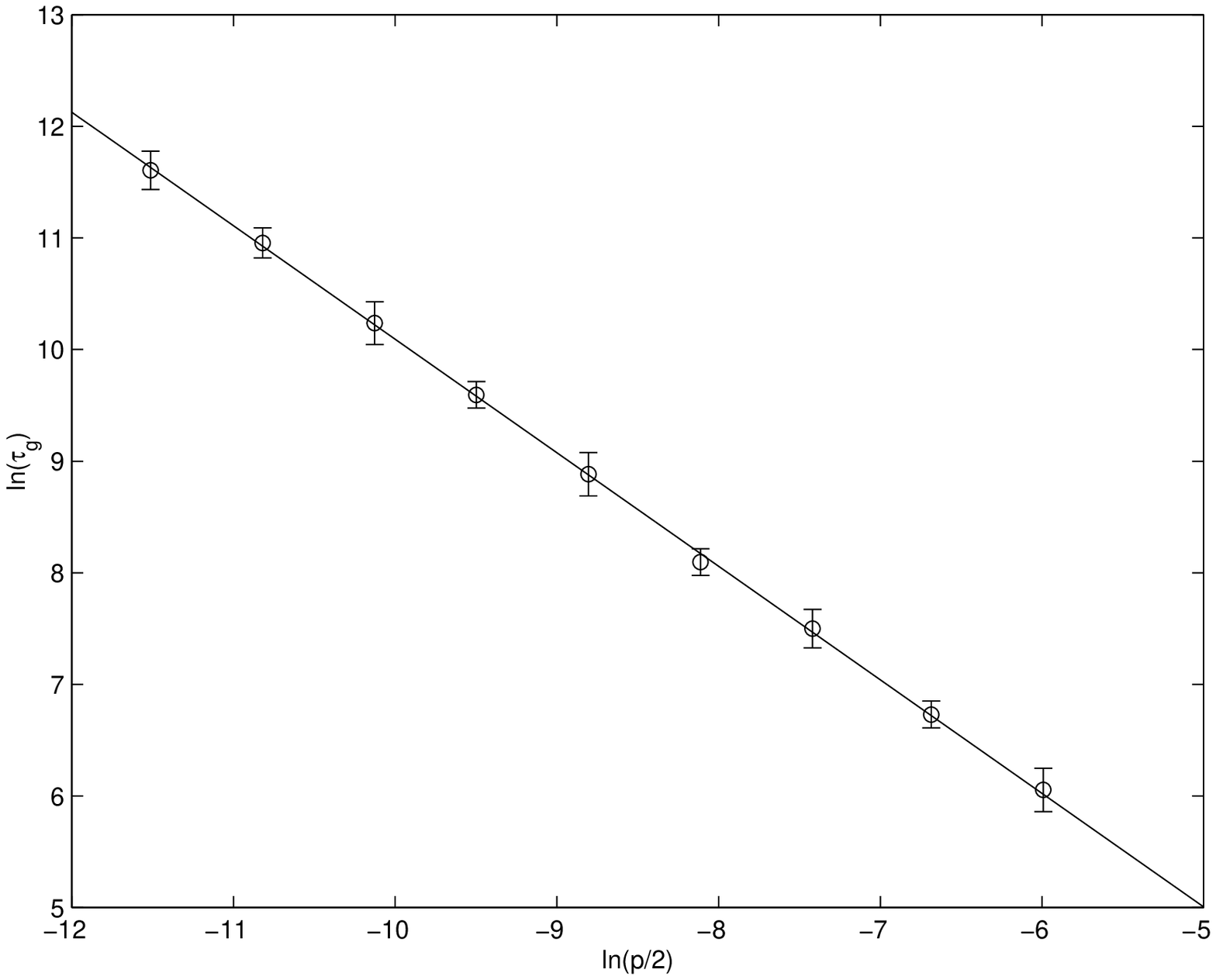} 

\vspace{0.1in}
\noindent
{\bf Figure 2.} Power law dependence of $\tau_g$ on $p$. 
Each data point shows the average of $\tau_g$ over 5 different runs with 
$s=100$ and the given $p$ value. The error bars correspond to one 
standard deviation.
The best fit line has slope $-1.02\pm 0.03$ and intercept $-0.08\pm 0.26$
which is consistent with the expected slope $-1$ and intercept $0$.\\\\\\

\noindent
\end{figure}

\begin{figure}
\epsfxsize=8cm
\epsfbox{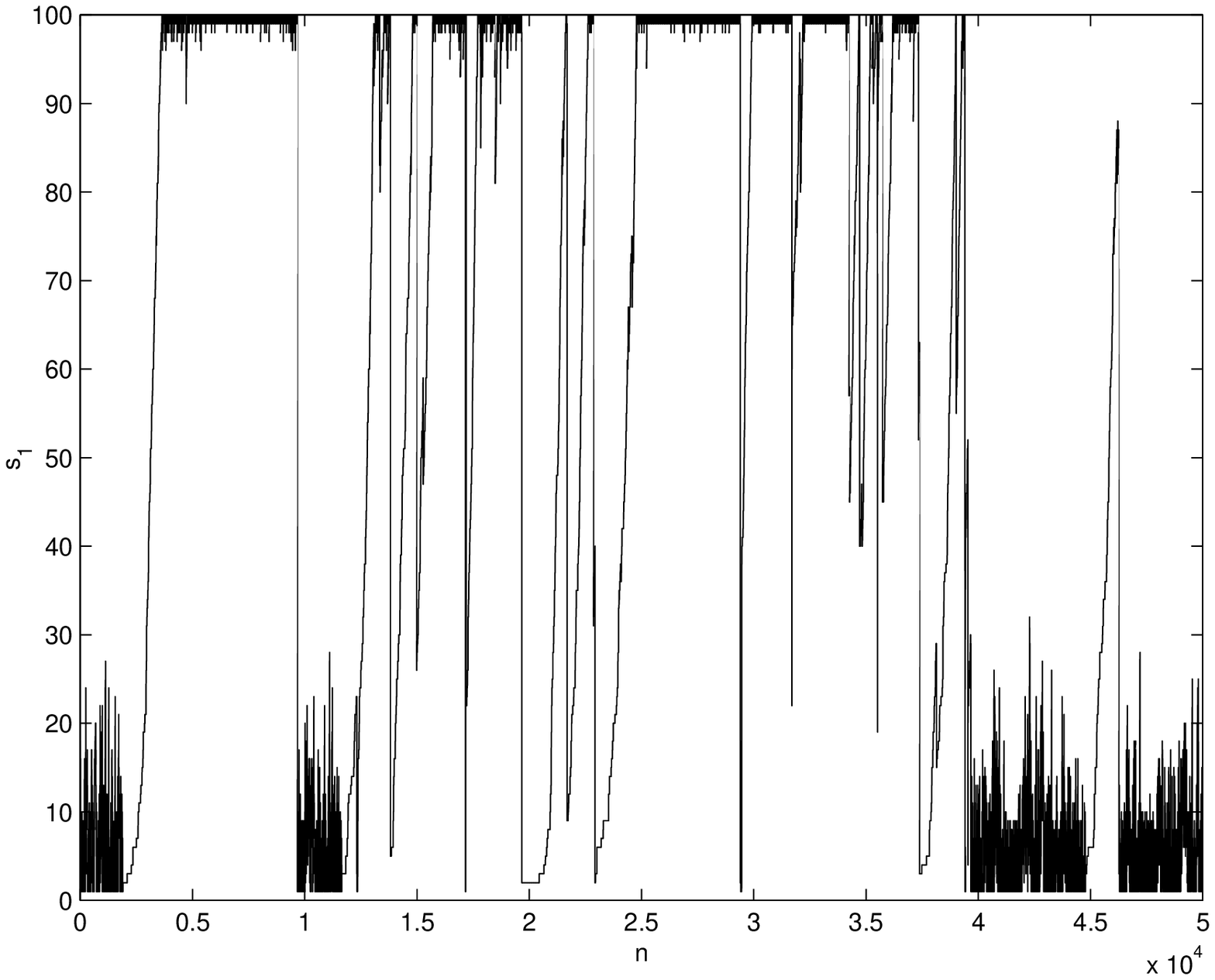} 

\vspace{0.1in}
\noindent
{\bf Figure 3.} The run of Figure 1a displayed over a much longer time scale.\\

\noindent
\end{figure}
\end{document}